\documentclass[prb,aps,nobalancelastpage,amssymb,twocolumn,citeautoscript]{revtex4}

\usepackage{graphicx}

\begin{document}
\preprint{PRB/Xu \textit{et al}.}

\title{Controllable spin-orbit coupling and its influence on the upper critical field in the chemically doped quasi-one-dimensional Nb$_2$PdS$_5$ superconductor}

\author{N. Zhou$^1$,  Xiaofeng Xu$^{1}$, J. R. Wang$^1$, J. H. Yang$^{1}$, Y. K. Li$^1$, Y. Guo$^1$, W. Z. Yang$^1$, C. Q. Niu$^1$, Bin Chen$^{2,1}$, Chao Cao$^1$, Jianhui Dai$^1$}
\affiliation{$^{1}$Department of Physics and Hangzhou Key Laboratory of Quantum Matters, Hangzhou Normal University, Hangzhou 310036, China\\
$^{2}$Department of Physics, University of Shanghai for Science $\&$ Tehcnology , Shanghai,
China\\}

\date{\today}

\begin{abstract}
By systematic chemical substitution of Pt and Ni in the newly-discovered superconductor
Nb$_2$PdS$_5$ ($T_c\sim$6 K), we study the evolution of its superconducting properties with doping,
focussing on the behavior of the upper critical field $H_{c2}$. In contrast to the previous results
of Se doping on S sites, superconductivity is found to be rather robust against the Pt and Ni
dopants on the one-dimensional Pd chains. Most strikingly, the reduced $H_{c2}$, i.e., the ratio of
$H_{c2}/T_c$, is seen to be significantly enhanced by the heavier Pt doping but suppressed in the
Ni-doped counterparts, distinct from the nearly constant value in the Se doped samples. Our
findings therefore suggest that the upper critical field of this system can be modified in a
tunable fashion by chemical doping on the Pd chains with elements of varying mass numbers. The
spin-orbit coupling on the Pd sites, by inference, should play an important role in the observed
superconductivity and on the large upper critical field beyond the Pauli pair-breaking field.
\end{abstract}

\maketitle

Whilst there exists much uncertainty in its profound microscopic details, a consensus begins to
emerge among physicists that the spin-orbit interactions may significantly change our established
picture in condensed matter physics and bring about an overwhelming amount of new fascinating
phenomena that would otherwise be highly
impossible\cite{Hasan10,Qi11,Goh12,Yang12,Taillefer13,Cava14,shimozawa14}. Notably, in topological
insulators, it is the spin-orbit coupling (SOC) that opens up a band gap in the bulk and gives rise
to the protected conducting surface states\cite{Hasan10,Qi11}. In addition, in some
non-centrosymmetric superconductors\cite{Yuan07,Zheng08,Yuan13,Yuan14}, the associated asymmetric
electrical field gradient may lead to an asymmetric SOC which may ultimately split the Fermi
surface (FS) into two segments with different spin structure. This allows the admixture of
spin-singlet and spin-triplet in the superconducting order parameter, with the ratio of which being
tunable by the strength of SOC, as nicely demonstrated in Li$_2$Pd$_3$B and
Li$_2$Pt$_3$B\cite{Yuan07,Zheng08}. Moreover, in the Pauli-limited superconductivity, SOC
counteracts the effect of the spin paramagnetism in limiting the upper critical field and leads to
a $H_{c2}$ significantly higher than the so-called weak-coupling Pauli limit,
$H_{c2}$=1.84$T_c$\cite{Zuo00}, as parameterized in the Werthamer-Helfand-Hohenberg (WHH)
theory.\cite{WHH66}

Recently, a new ternary compound Nb$_2$PdS$_5$ with one-dimensional Pd chains has been reported to
be superconducting below $T_c$$\sim$ 6 K\cite{Zhang13,Zhang132,Yu13,Niu13,Khim13,Takagi14,Awana14}.
Remarkably, its upper critical magnetic field $H_{c2}$ along the chains was observed to exceed the
Pauli paramagnetic limit by a factor of 3\cite{Zhang13,Niu13}. Tentatively, this large upper
critical field can be ascribed to the strong-coupling theory\cite{Carbotte90,Mizukami11},
spin-triplet pairing\cite{Raghu10,Lee00,Lee01,Xu09,Mercure12}, strong spin-orbit
interaction\cite{WHH66,Niu13,Owen07} or the multi-band effect\cite{Hunte08,Xu13}. However, the
previous calorimetric study seemingly ruled out the strong-coupling and spin-triplet pairing as its
origin\cite{Niu13}. Since this superconductor involves heavy element Pd, SOC ought to be large,
recalling that SOC is proportional to $Z^4$, where $Z$ is the atomic mass number\cite{shiyan13}.
Nevertheless, the role of the SOC on its superconducting properties, especially on its high upper
critical field has not been resolved thus far.

In this paper, we use the chemical substitution as a probe to study the role of SOC on the upper
critical field of Nb$_2$PdS$_5$. By substituting Pd with the heavier Pt, it is shown that the ratio
of $H_{c2}/T_c$ is significantly enhanced, while in the samples with lighter Ni doping, this ratio
is lessened. Compared to the Se doped series\cite{Niu13}, in which the ratio is very weakly
affected, our findings seem to suggest the role of the SOC on the large upper critical field in
this system. In the end of the paper, we also discuss the other possibilites, such as the
charge-density-wave (CDW) fluctuations, which are often associated with the one-dimensional
electrons, in the elevated $H_{c2}$ in this quasi-one-dimensional compound.

\begin{figure*}
\includegraphics[width=15cm,keepaspectratio=true]{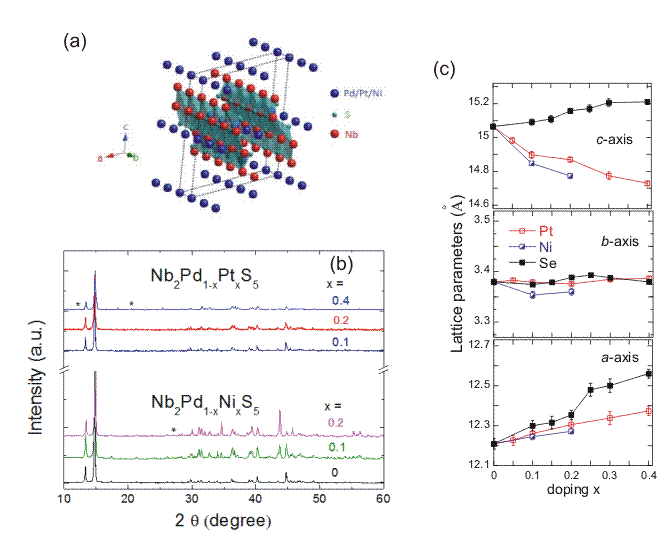}
\caption{(Color online) (a) The crystallographic structure of Nb$_2$(Pd$_{1-x}$Pt$_x$)S$_5$. The
one-dimensional Pd chains are extended along the $b$-axis. The unit cell is indicated by the thin
dotted line. (b) shows the powder XRD patterns for Pt and Ni doped samples studied here ($x$=0.05
and 0.3 of Pt-doped samples are not included for clarity), with the asterisks marking the possible
impurity phases. (c) The resultant lattice parameters extracted from the Rietveld refinement, along
with those for the Se doped samples for comparison.} \label{Fig1}
\end{figure*}

Nb$_2$(Pd$_{1-x}$Pt$_x$)S$_5$ ($x$=0, 0.05, 0.1, 0.2, 0.3, 0.4) and Nb$_2$(Pd$_{1-x}$Ni$_x$)S$_5$
($x$=0.1, 0.2) Polycrystalline samples were synthesized by a solid state reaction in
vacuum\cite{Keszler85}. The details of the sample growth procedure were described
elsewhere\cite{Niu13}. Good crystallinity of the as-grown samples was confirmed by a X-ray powder
diffractometer. Lattice parameters were obtained by Rietveld refinements. The schematic of
crystallographic structure is shown in Figure 1(a), where the one-dimensional Pd/Pt/Ni chains are
oriented along the $b$-axis. (Magneto-)resistance was measured by a standard four-probe lock-in
technique in an applied field up to 9 tesla in Quantum Design PPMS. Specific heat measurements were
also performed with this facility. The temperature dependence of the dc magnetization was done in
MPMS-7 system, with both zero-field-cooling (ZFC) and field-cooling (FC) modes being employed to
probe the superconducting transitions.

The X-ray diffraction (XRD) shown in Fig. 1(b) can be well fit to the crystallographic structure
depicted in Fig. 1(a) with monoclinic space group $C$2/$m$\cite{Zhang13}. Only a small trace of
impurity phase, marked by the asterisk, was detected. The lattice parameters were extracted and
plotted with those of Se doped samples from Ref. [17] for comparison, as shown in Fig. 1(c). It is
noted that all these three dopants increase the $a$-axis lattice by $>$1$\%$ while the $b$-axis
length remains roughly constant. Remarkably, whilst the Se doping expands the $c$-axis lattice
notably, the incorporated Pt and Ni ions significantly shrink the lattice along this direction.

\begin{figure}
\includegraphics[width=6cm,keepaspectratio=true]{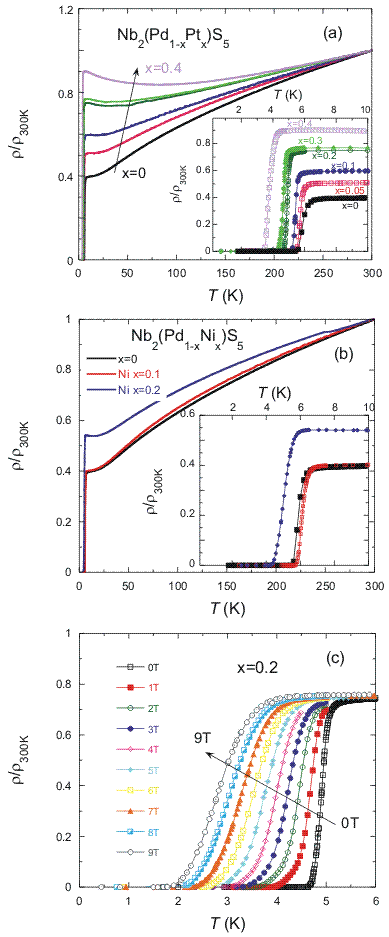}
\caption{(Color online) (a) and (b) depict the zero-field resistivity for Pt-doped and Ni-doped
samples studied in this work, respectively. Note all curves are renormalized to their room
temperature values for the purpose of clarity. The insets blow up the superconducting transitions.
(c) The temperature sweeps at some fixed-fields for $x$=0.2 Pt-doped sample as an example. }
\label{Fig2}
\end{figure}

Figure 2(a) and (b) show respectively the zero-field resistivity of Pt- and Ni-doped samples
studied here, divided by their individual room temperature values for clarity. Their respective
insets zoom in the low temperature superconducting transitions below 10 K. It should be noted that
the residual resistivity ratio systematically decreases with increasing Pt- and Ni-dopings. For
Nb$_2$(Pd$_{1-x}$Pt$_x$)S$_5$, once $x\geq$0.1, a well-defined resistivity minimum appears below
$T_{\texttt{min}}$ in the normal state, similar to the Se doped series\cite{Niu13}. For Ni-doped
case, it is found that 10$\%$ adopted Ni ions only change the reduced resistivity curve slightly
compared to the parent compound, and no resistivity minimum appears until $x$=0.2. We attribute the
resistivity dips to the possible disorder-induced localization effect. The temperature dependence
of the upper critical field $H_{c2}(T)$ for each sample was then determined by the temperature
sweeps at fixed fields, exemplified in Fig. 2(c) for Pt-doped x=0.2 sample. We shall return to this
point later.

\begin{figure*}
\includegraphics[width=8cm,keepaspectratio=true]{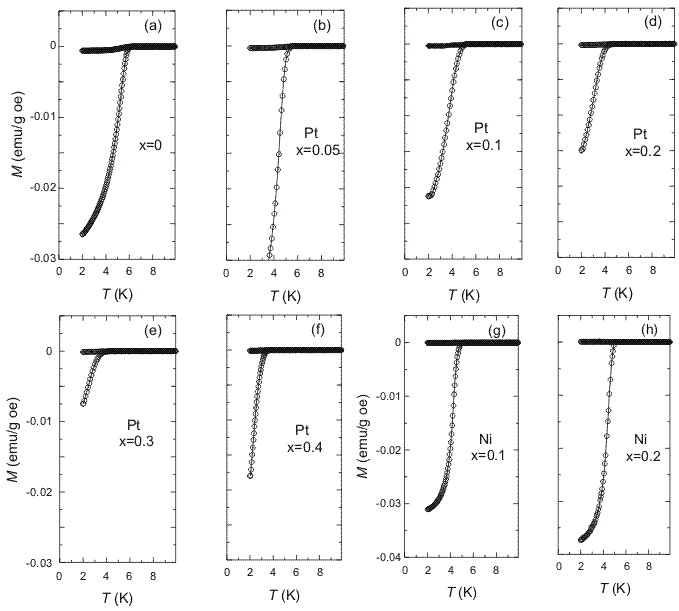}
\caption{(Color online) The magnetizations below $T_c$ with ZFC and FC modes. The data were taken
under a 10 Oe magnetic field. } \label{Fig4}
\end{figure*}

The bulk nature of the superconductivity for all doping samples was confirmed by the magnetization
measurements, as given in Fig. 3. The large diamagnetic signals were clearly seen for all samples.
The total heat capacity of all samples below 10 K was given in order in Fig. 4. The heat capacity
anomalies associated with the superconducting transition were clearly seen in all dopings studied.
However, as the measurements were only performed down to 2 K (except for the parent compound which
is down to 0.5 K), and for some doping levels, there are still significant amount of
non-superconducting volumes, this precludes us from the accurate determination of electronic
$\gamma$ term, condensation energy and superconducting coupling strength so on, as a function of
doping. Higher sample quality and extra low-$T$ measurements are required in the future to reveal
the possible correlations between all these physical quantities.

As exemplified in Fig. 2(c), the resultant $H_{c2}(T)$ for both series of samples was summarized in
Fig. 5 (a) and (c) respectively, together with their corresponding WHH fits. The applicability of
the WHH fits was assured as the fitting was demonstrated to capture the whole $H_{c2}(T)$ profile
for the single crystal Nb$_2$PdS$_5$ up to 40 tesla in Ref. [17]. $T_c$ as well as the as-drawn
$H_{c2}(0)$ (i.e., $H_{c2}$ at $T$=0 K) was given in Fig. 5(b) and (d) correspondingly. For
Nb$_2$(Pd$_{1-x}$Pt$_x$)S$_5$ samples, it is remarkable that, although $T_c$ monotonically
decreases with Pt content, $H_{c2}$ initially goes up with the Pt substitution before being
suppressed finally. This finding is in sharp contrast to the behaviors observed in the Se doped
study, where both $T_c$ and $H_{c2}$ were simultaneously suppressed by doping\cite{Niu13}. As for
the Ni-doped samples, $T_c$ slightly goes up for $x$=0.1 and decreases with further doping.

\begin{figure*}
\includegraphics[width=8cm,keepaspectratio=true]{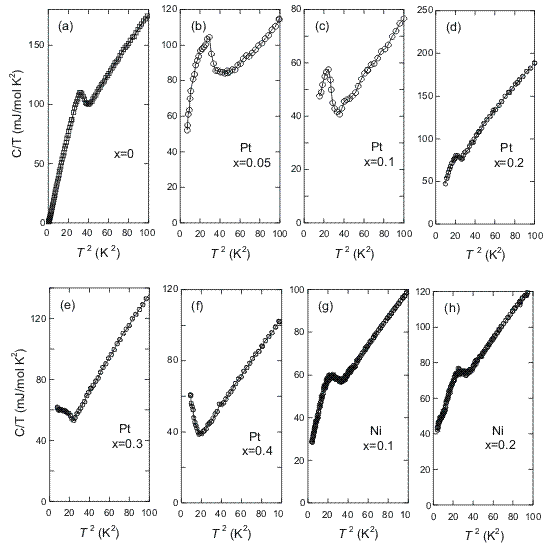}
\caption{(Color online) The specific heat data for all samples in zero magnetic field, plotted as
$C/T$ vs $T^2$, to demonstrate the anomalies associated with the superconducting transition.}
\label{Fig4}
\end{figure*}

The $T$ vs doping diagram, derived collectively from all three series of samples, are presented in
Fig. 6(a). Interestingly, $T_c$ and $T_{\texttt{min}}$ show the anti-correlation with the doping.
While $T_{\texttt{min}}$ increases with doping, $T_c$ goes down, with a much sharper suppression in
the Se doped samples. This indicates that the superconductivity is rather robust against the
isovalent chemical doping on the one-dimensional Pd chains. This counter-intuitive robustness of
superconductivity against the impurities sited on the one-dimensional chains is surprising because
the impurities in the one-dimensional chains usually serve as strong back-scatters and electrons
are apt to be localized as a result, as seen in the other quasi-one-dimensional
PrBa$_2$Cu$_4$O$_8$\cite{Narduzzo07,Rad07}.

\begin{figure*}
\includegraphics[width=10cm,keepaspectratio=true]{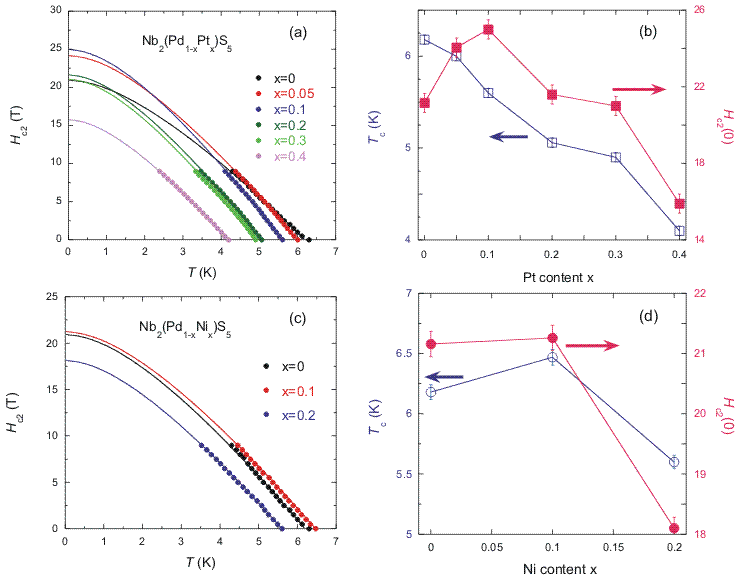}
\caption{(Color online) (a) and (c) give the temperature dependence of $H_{c2}$ for
Nb$_2$(Pd$_{1-x}$Pt$_x$)S$_5$ and Nb$_2$(Pd$_{1-x}$Ni$_x$)S$_5$ respectively, extracted from the
temperature sweeps at constant fields, determined by 90$\%$ of the normal state resistivity. The
solid lines represent the WHH fittings. $T_c$ and $H_{c2}$(0) (zero-temperature upper critical
field from the fit) are plotted in panel (b) and (d) accordingly. } \label{Fig2}
\end{figure*}

Fig. 6(b) contains the key finding of this study, namely, the strength of the upper critical field,
measured by the ratio $H_{c2}/T_c$, is significantly enhanced by the Pt doping in Nb$_2$PdS$_5$
system, yet reduced in the Ni-doped ones. Unlike the nearly constant value of $H_{c2}/T_c$ in
Nb$_2$Pd(S$_{1-x}$Se$_x$)$_5$\cite{Niu13}, the ratio goes up substantially by the initial Pt
substitution and decreases with further content. It is noted that in $x$=0.4 Pt-doped sample, this
ratio is considerably suppressed. The origin for this suppression is unknown to us but it is
presumably related to the impurity at such a high doping level, evidenced from XRD and heat
capacity measurements. Importantly, the ratio in samples doped with Pt is larger than in samples
doped with the same amount of Se at \textit{all} doping levels. This is consistent with the
argument that the large upper critical field in Nb$_2$PdS$_5$ is due to the large spin-orbit
coupling inherent in the heavy Pd elements\cite{Khim13,Takagi14}. When the heavier Pt ions are
incorporated, the SOC is further enhanced, therefore it gives rise to a larger upper critical field
(and a larger $H_{c2}/T_c$). This simple argument is again corroborated by the lighter Ni doping.
The results on Ni-doped samples analyzed on the same footing clearly point to a smaller
$H_{c2}/T_c$ ratio, albeit in the less pronounced manner, as shown in Fig. 6(b).

Finally, let us consider other possible origins for the observed contrasting doping dependence of
$H_{c2}/T_c$ ratio. It is tempting to argue that the enhanced $H_{c2}/T_c$ ratio in
Nb$_2$(Pd$_{1-x}$Pt$_x$)S$_5$ is due to the internal pressure effect induced by the chemical
doping. As seen from Fig. 1(c), the $c$-axis decreased substantially with Pt concentration,
corresponding to the positive pressure along the $c$-axis. On the contrary, the $c$-axis in
Nb$_2$Pd(S$_{1-x}$Se$_x$)$_5$ increases with doping. One would expect the reduced $H_{c2}/T_c$
ratio by the Se doping due to the negative pressure. However, this is \textit{not} seen and in
fact, the ratio is barely modified by the Se doping (Note that it even slightly increases for
$x$=0.3 and 0.4). The same doping dependence of lattice parameters but the opposite $H_{c2}/T_c$
tendency in the Pt- and Ni-doped samples also rules out the pressure effect as the origin of their
contrasting $H_{c2}/T_c$ slope.

On the other hand, it is well known that in many low-dimensional materials, Fermi surfaces are
susceptible to the CDW instability below a critical temperature $T_p$. According to
theory\cite{Gabovich04}, in superconductors with CDW correlations, the paramagnetic limit of
$H_{c2}$ can be greatly enhanced, depending on the relative temperature scale of $T_c$ and $T_p$,
as well as the nested FS portion. Regarding Nb$_2$PdS$_5$ system, whereas the long-range CDW
formation is not evidently seen in the resistivity, any CDW fluctuations associated with the
one-dimensional FS may also promote its $H_{c2}$ to a value much higher than the Pauli
limit\cite{Ohmichi99}.

According to band structure calculations\cite{Zhang13,Khim13}, the FS of Nb$_2$PdS$_5$ consists of
a set of electron-like flat sheets, closed pockets and hole-like corrugated cylinders, mainly from
the $d$-orbitals of Nb and Pd atoms. How the substituent elements Pt, Ni and Se change the band
structures (including those far from the Fermi level) is unknown to us. It would be challenging for
band structure calculations to explain simultaneously the doping dependence of \textit{all}
quantities, $T_c$, $H_{c2}$ and $H_{c2}/T_c$, revealed in this study. Moreover, it is also likely
that the incorporation of dopants changes the effective dimensionality of the compound, which may
affect $H_{c2}$ accordingly.

\begin{figure}
\includegraphics[width=6cm,keepaspectratio=true]{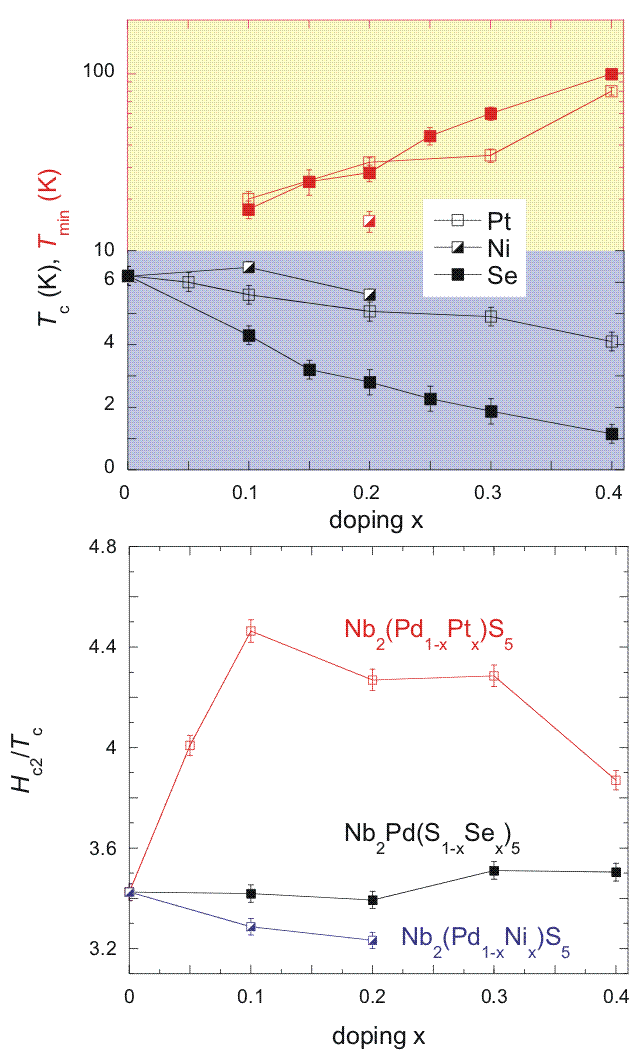}
\caption{(Color online)  Top panel: The doping dependence of $T_c$ and $T_{\texttt{min}}$ as a
function of doping. Data for Se doped samples are also included. Note the log-scale for
$T_{\texttt{min}}$. Bottom panel: The doping dependence of the ratio $H_{c2}/T_c$ for Pt and Ni
substituted samples, plotted together with those for Se-doped ones.} \label{Fig2}
\end{figure}

In conclusion, we have studied the effect of Pt and Ni substitutions on the one-dimensional Pd
chains in Nb$_2$PdS$_5$ superconductor. By comparison with the previous Se doping in this system,
we revealed the significant differences in their structural and physical properties, in particular
in the ratio of $H_{c2}/T_c$. The contrasting behaviors of $H_{c2}/T_c$ in these three series have
been tentatively attributed to the differences in their strength of SOC. Our study suggest that SOC
should play a significant role on the large upper critical field beyond the Pauli paramagnetic
limit in this system, although other factors, such as the the multi-band effects\cite{Zhang13} and
CDW fluctuations or superstructure\cite{Gabovich04}, may be also at play here.

The authors would like to thank N. E. Hussey, C. M. J. Andrew, C. Lester, E. A. Yelland, A. F.
Bangura, Xiaoyong Feng, Wenhe Jiao, Huifei Zhai, Guanghan Cao, Xiaofeng Jin for stimulating
discussions and P. Biswas for collaborative support. This work is sponsored by the National Key
Basic Research Program of China (Grant No. 2014CB648400) and by NSFC (Grant No. 11474080, 11104051,
11104053). X.X. would also like to acknowledge the financial support from the Distinguished Young
Scientist Funds of Zhejiang Province (LR14A040001).

\bibliography{NbPdPtS}

\begin{thebibliography}{27}

\bibitem{Hasan10} M. Z. Hasan, C. L. Kane, Rev. Mod. Phys. {\bf 82}, 3045 (2010).
\bibitem{Qi11} Xiao-Liang Qi, Shou-Cheng Zhang, Rev. Mod. Phys. {\bf 83}, 1057 (2011).
\bibitem{Goh12} S. K. Goh, Y. Mizukami, H. Shishido, D. Watanabe, S. Yasumoto, M. Shimozawa, M. Yamashita, T. Terashima, Y. Yanase, T. Shibauchi, A. I. Buzdin, and Y. Matsuda, Phys. Rev. Lett. {\bf 109}, 157006 (2012).
\bibitem{Yang12} J. J. Yang, Y. J. Choi, Y. S. Oh, A. Hogan, Y. Horibe, K. Kim, B. I. Min, and S. W. Cheong, Phys. Rev. Lett. {\bf 108}, 116402 (2012).
\bibitem{Taillefer13} F. F. Tafti, Takenori Fujii, A. Juneau-Fecteau, S. Rene de Cotret, N. Doiron-Leyraud, Atsushi Asamitsu, and Louis Taillefer, Phys. Rev. B {\bf 87}, 184504 (2013).
\bibitem{Cava14} M. N. Ali, Q.D. Gibson, T. Klimczuk, R. J. Cava, Phys. Rev. B {\bf 89}, 020505 (2014).
\bibitem{shimozawa14} M. Shimozawa, S. K. Goh, R. Endo, R. Kobayashi, T. Watashige, Y. Mizukami, H. Ikeda, H. Shishido, Y. Yanase, T. Terashima, T. Shibauchi, and Y. Matsuda, Phys. Rev. Lett. {\bf 112}, 156404 (2014).
\bibitem{Yuan07} H. Q. Yuan, D. F. Agterberg, N. Hayashi, P. Badica, D. Vandervelde, K. Togano, M. Sigrist, and M. B. Salamon, Phys. Rev. Lett. {\bf 97}, 017006 (2006).
\bibitem{Zheng08} M. Nishiyama, Y. Inada, and G. Q. Zheng, Phys. Rev. Lett. {\bf 98}, 047002 (2007).
\bibitem{Yuan13} J. Chen, L. Jiao, J. L. Zhang, Y. Chen, L. Yang, M. Nicklas, F. Steglich, and H. Q. Yuan, Phys. Rev. B {\bf 88}, 144510 (2013).
\bibitem{Yuan14} L. Jiao, J. L. Zhang, Y. Chen, Z. F. Weng, Y. M. Shao, J. Y. Feng, X. Lu, B. Joshi, A. Thamizhavel, S. Ramakrishnan, and H. Q. Yuan, Phys. Rev. B {\bf 89}, 060507 (2014).
\bibitem{Zuo00} F. Zuo, J. S. Brooks, R. H. Mckenzie, J. A. Schlueter, and J. M. Williams, Phys. Rev. B {\bf 61}, 750 (2000).
\bibitem{WHH66} N. R. Werthamer, E. Helfand, P. C. Hohenberg Phys. Rev. {\bf 147}, 295 (1966).
\bibitem{Zhang13} Q. Zhang, G. Li, D. Rhodes, A. Kiswandhi, T. Besara, B. Zeng, J. Sun, T. Siegrist, M. D. Johannes and L. Balicas, \textit{Sci. Rep.} {\bf 3}, 1446 (2013).
\bibitem{Zhang132} Q. Zhang, D. Rhodes, B. Zeng, T. Besara, T. Siegrist, M. D. Johannes, L. Balicas, Phys. Rev. B {\bf 88}, 024508 (2013).
\bibitem{Yu13} H. Y. Yu, M. Zuo, L. Zhang, S. Tan, C. J. Zhang, and Y. H. Zhang, J. Am. Chem. Soc. {\bf 135}, 12987 (2013).
\bibitem{Niu13} C. Q. Niu, J. H. Yang, Y. K. Li , Bin Chen, N. Zhou, J. Chen, L. L. Jiang, B. Chen, X. X. Yang, Chao Cao, Jianhui Dai, and Xiaofeng Xu, Phys. Rev. B {\bf 88}, 104507 (2013).
\bibitem{Khim13} S. Khim, B. Lee, K. Y. Choi, B. G. Jeon, D. H. Jang, D. Patil, S. Patil, R. Kim, E. S. Choi, S. Lee, J. Yu and K. H. Kim, New J. Phys. {\bf 15}, 123031 (2013).
\bibitem{Takagi14} Y. F. Lu, T. Takayama, A. F. Bangura, Y. Katsura, D. Hashizume, G. Li, and H. Takagi, J. Phys. Soc. Jpn. {\bf 83}, 023702 (2014).
\bibitem{Awana14} R. Jha, B. Tiwari, P. Rani, V.P.S. Awana, arXiv:1312.0425
\bibitem{Carbotte90} J. P. Carbotte, Rev. Mod. Phys. {\bf 62}, 1027 (1990).
\bibitem{Mizukami11} Y. Mizukami, H. Shishido, T. Shibauchi, M. Shimozawa, S. Yasumoto, D. Watanabe, M. Yamashita, H. Ikeda, T. Terashima, H. Kontani and Y. Matsuda, Nat. Phys. {\bf 7}, 849 (2011).
\bibitem{Raghu10} S. Raghu, A. Kapitulnik, and S. A. Kivelson, Phys. Rev. Lett. {\bf 105}, 136401 (2010).
\bibitem{Lee00} I. J. Lee, P. M. Chaikin, and M. J. Naughton, Phys. Rev. B {\bf 62}, 14669 (2000).
\bibitem{Lee01} I. J. Lee, S. E. Brown, W. G. Clark, M. J. Strouse, M. J. Naughton, W. Kang, and P. M. Chaikin, Phys. Rev. Lett. {\bf 88}, 017004 (2001).
\bibitem{Xu09} Xiaofeng Xu, A. F. Bangura, J. G. Analytis, J. D. Fletcher, M. M. J. French, N. Shannon, J. He, S. Zhang, D. Mandrus, R. Jin, and N. E. Hussey, Phys. Rev. Lett. {\bf 102}, 206602 (2009).
\bibitem{Mercure12} J.-F. Mercure, A. F. Bangura, Xiaofeng Xu, N. Wakeham, A. Carrington, P. Walmsley, M. Greenblatt, and N. E. Hussey, Phys. Rev. Lett. {\bf 108}, 187003 (2012).
\bibitem{Owen07} O. J. Taylor, A. Carrington, J. A. Schlueter, Phys. Rev. Lett. {\bf 99}, 057001 (2007).
\bibitem{Hunte08} F. Hunte, J. Jaroszynski, A. Gurevich, D. C. Larbalestier, R. Jin, A. S. Sefat, M. A. McGuire, B. C. Sales, D. K. Christen, D. Mandrus, Nature {\bf 453}, 903 (2008).
\bibitem{Xu13} Xiaofeng Xu, B. Chen, W. H. Jiao, Bin Chen, C. Q. Niu, Y. K. Li , J. H. Yang, A. F. Bangura, Q. L. Ye, C. Cao, J. H. Dai, Guanghan Cao, and N. E. Hussey, Phys. Rev. B {\bf 87}, 224507 (2013).
\bibitem{shiyan13} S. Y. Zhou, X.L. Li, B.Y. Pan, X. Qiu, J. Pan, X.C. Hong, Z. Zhang, A.F. Fang, N. L. Wang, and S. Y. Li, EPL, {\bf 104} 27010 (2013).
\bibitem{Keszler85} D. A. Keszler, J. A. Ibers, Maoyu Shang and Jiaxi Lu, J. Solid State Chem. {\bf 57}, 68 (1985).
\bibitem{Narduzzo07} A. Narduzzo, A. Enayati-Rad, S. Horii, and N. E. Hussey, Phys. Rev. Lett. {\bf 98}, 146601 (2007).
\bibitem{Rad07} A. Enayati-Rad, A. Narduzzo, F. Rullier-Albenque, S. Horii, and N. E. Hussey, Phys. Rev. Lett. {\bf 99}, 136402 (2007).
\bibitem{Gabovich04} A. Gabovich, A. I. Voitenko, and T. Ekino, J. Phys.:Condens. Matter {\bf 16}, 3681 (2004) and references therein.
\bibitem{Ohmichi99} E. Ohmichi, T. Ishiguro, T. Sakon, T. Sasaki, M. Motokawa, R. B. Lyubovskii, and R. N. Lyubovskaya, J. Supercond. {\bf 12}, 505 (1999).

\end{thebibliography}


\end{document}